\documentstyle[prb,aps,epsf]{revtex}
\begin{document}

%2
\flushbottom
%2
\twocolumn[\hsize\textwidth\columnwidth\hsize\csname 
%2
@twocolumnfalse\endcsname 

\title{Prediction of Room Temperature High Thermoelectric Performance in n-type
La(Ru$_{1-x}$Rh$_x$)$_4$Sb$_{12}$} 

\author{Marco Fornari}
\address{C.S.I. George
Mason University, Fairfax, Virginia 22030-4444 and Naval Research Laboratory,
Washington D.C. 20375-5345}   
\author{David J. Singh}
\address{Center for Computational Material Science, Naval Research Laboratory,
Washington D.C. 20375-5345}

\date{\today}
\maketitle

\begin{abstract}
%1\setlength{\baselineskip}{0.7cm}
First principles calculations are used to investigate the band
structure and the transport related properties of unfilled and filled
4$d$ skutterudite antimonides. The calculations show that, while
RhSb$_3$ and p-type  La(Rh,Ru)$_4$Sb$_{12}$ are unfavorable for
thermoelectric application, n-type La(Rh,Ru)$_4$Sb$_{12}$ is very
likely a high figure of merit thermoelectric material in the important
temperature range 150-300 K.
\end{abstract}

%2
]
%1\newpage

% Introduction
The widespread application of thermoelectric (TE) coolers has been
hindered by the fact that these devices have limited efficiencies
related to the lack of sufficiently high performance materials to use
in the active elements. The TE performance of a material
is characterized by a dimensionless figure of merit $\mbox{ZT}= \sigma S^2
\mbox{T}/(k_L+k_e)$  where $\sigma$ and $k_e$ are the
electrical and thermal conductivity, $S$ is the Seebeck coefficient
and $k_L$ is the lattice thermal conductivity. Almost all TE coolers
are based on alloys of Bi$_2$Te$_3$, which has maximum ZT values
slightly exceeding 1.0 around room temperature, but substantially
decreasing performance as temperature is reduced.\cite{m-ssp} 

Recently, there has been a renewed effort to find better materials in
the temperature range 100-300 K with a primary focus on novel
systems. This effort
has been partially successful in that three new high ZT compounds have been
discovered: $\beta$-Zn$_4$Sb$_3$,\cite{cfb-znsb-prb} and two filled
skutterudites CeFe$_4$Sb$_{12}$,\cite{fbcmm-ict} and
La(Fe,Co)$_4$Sb$_{12}$.\cite{smw-s} While these materials all have
their maximum ZT values substantially above room temperature and are
not applicable to cooling applications, considerable understanding,
particularly of skutterudites, has emerged leading to the hope
that even higher performance compositions may be found and perhaps
that lower temperature operation may be possible. Here we report first
principles results that when combined with existing knowledge of
skutterudites materials, point to a specific composition with a strong
potential for high values of ZT (in excess of 1.0) over a wide
technologically important temperature range from 150-300 K. 

The conventional cubic cell of a binary skutterudite, for instance
CoSb$_3$, contains eight molecular units (space group
$Im\overline{3}$). The metal can be also
Rh or Ir, and Sb an other pnictogen element (P, As). 
The metal atoms form a cubic sub-lattice partially filled by
six almost square pnictogen rings oriented according to the cubic
directions. This results in six bonds connecting one Co with the
six surrounding Sb  whereas every pnictogen forms two strong
covalent bond with other Sb and two partially ionic bonds with Co. 
The high values of ZT, with a rare earth filling the empty
octans of the cubic cell, rests on two
important effects: a low glass-like $k_L$ due to the effect of the
filling atom vibrations,\cite{smckt-prb} and on high
power factors $\sigma S^2$ related to  the electronic features. 
In La(Fe,Co)$_4$Sb$_{12}$, in particular, La-filling has the important
effect to decrease $k_L$ about one order of magnitude,\cite{smw-s} with respect
binary CoSb$_3$ and to lower the
top of the valence band to near a low-energy high 
effective mass band. This is responsible for
the high thermopower in p-type materials.\cite{sm-rprb} 

In this letter, we investigate by means of first principles calculations the
electronic properties of skutterudites based on RhSb$_3$ and find that
La(Ru,Rh)$_4$Sb$_{12}$ is promising for n-type TE application. 
Although the thermal conductivity of RhSb$_3$ is not glass-like,\cite{fcb-aip} 
we can expect  that filling with La will yield a low $k_L$
material because the volume available to La's vibration allows
considerable amplitudes and a clear relation between filling and $k_L$
in skutterudites is known experimentally. This conjecture is supported
by the experimental 
finding that $k_L$ in LaFe$_4$Sb$_{12}$  is lower than 20
mWcm$^{-1}$K$^{-1}$. Based on experimental trends of $k_L$ vs. void
size in skutterudites we expect $k_L$ for La(Ru,Rh)$_4$Sb$_{12}$ to be
in the range 10-15 mWcm$^{-1}$K$^{-1}$.

Our electronic structure calculations were performed using a well
converged ($R_{KM} =$ 7.0, $R_{MT} =$ 2.50, 2.25, 2.25, 2.20 for La,
Ru, Rh, Sb respectively) linearized augmented plane wave (LAPW)
method, \cite{s-lapw} 
including $s$, $p$, and $d$ local orbital extensions, \cite{s-prb} in
the framework of local 
density functional theory (Hedin-Lundqvist parameterization). 
A fully relativistic calculation was used for core states whereas the
valence states were done in a scalar relativistic scheme.
The Brillouin zone (BZ) sampling in the self consistent calculation is done
by means of a (4,4,4) special point grid whereas the density of states,
$N(\varepsilon)$, and related quantities are integrated on 295 {\bf k}
points tetrahedral mesh. 

We used the experimental lattice constant for LaRu$_4$ Sb$_{12}$ ($a_L =$
9.26 \AA); the two symmetry unconstrained structural parameters
related to the Sb position were determined by means of a best fit to
a quadratic curve of total energy 
calculations exploring 2-parameter surface. We obtain $x =$ 0.1583 and
$y =$ 0.3412 in the notation of Ref.\onlinecite{hand}. These structural
parameters were used also for the calculation on La(Ru,Rh)$_4$Sb$_{12}$
performed in virtual crystal approximation (VCA), realized in LAPW
adding one electron in valence and increasing the atomic number of the
transition metal of one fourth. As discussed below, excellent rigid
band behavior was found between LaRu$_4$Sb$_{12}$ and VCA
La(Ru,Rh)$_4$Sb$_{12}$ 
supporting the validity of VCA, {\it a posteriori}, and implying only weak
scattering by alloy disorder.

\begin{figure}%%%%%%%%%%% fig 1
\epsfbox{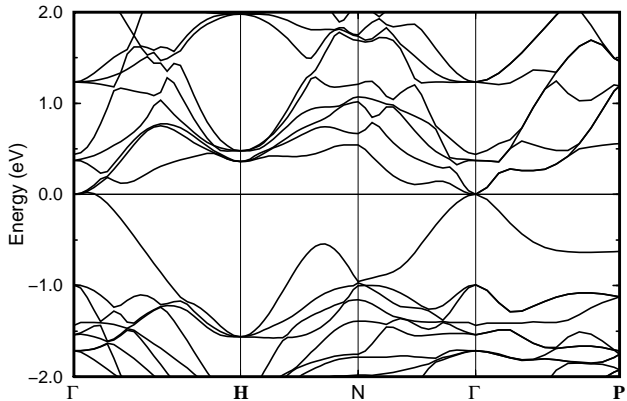}
\caption{Scalar relativistic band structure for RhSb$_3$ in an energy
window centered at the valence band top. Notice the zero-gap character
due to 3-fold degenerate band at BZ center.}
\label{BS1}
\end{figure}

RhSb$_3$, whose experimental structural parameters are $a_L =$
9.23 \AA\ and $(x,y)=$ (.1517, .3420), is a zero-gap semiconductor
because the top of the valence band at BZ center belongs to a 3-fold
degenerate representation (see Fig.\ \ref{BS1}). Anyway the
highest occupied band shows strong Sb-$p$ component that might
interact, after filling, with La-$f$ resonance,\cite{sm-rprb} as
happens both in cobalt phosphide and antimonide. 

The band structure (BS) of the corresponding filled material,
LaRu$_4$Sb$_{12}$, is drastically different from binary
RhSb$_3$ because the effect of La is to introduce $f$-resonance in
conduction band (CB) and to make available extra-electrons that modify
the electronic structure. This has been proved studying the BS of
hypothetical YRu$_4$Sb$_{12}$. The different arrangement of the bands near
$\epsilon_F$ in RhSb$_3$ with respect of CoSb$_3$, as in
Ref.\onlinecite{sp-prb},  could be strongly related to the different
electro-negativity of Rh (Pauling electro-negativity: e$_P=$ 2.28)
and Co (e$_P=$ 1.88) with respect of Sb (e$_P=$ 2.05) inverting
the ionicity of bonds.     

Undoped LaRu$_4$Sb$_{12}$ is a metal but its BS is very promising
for n-type TE application if alloying can be done that brings
$\epsilon_F$ to the gap without changing the main features of the
heavy multi-valley structure above the indirect gap in the CB;
p-type TE is not favorable at reasonable doping because of the lack
of heavy hole bands near the Fermi level.

\begin{figure}%%%%%%%%%%% fig 2
\epsfbox{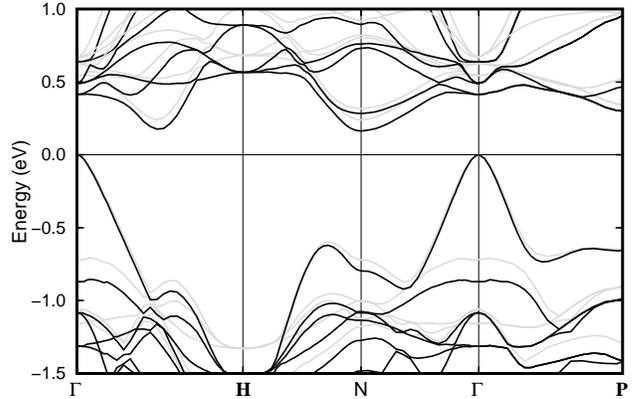}
\caption{Comparison between the scalar relativistic bands for
LaRu$_4$Sb$_{12}$ (gray line) and VCA La(Ru$_{.75}$Rh$_{.25}$)$_4$Sb$_{12}$
(black). The energy zero is fixed at the respective tops of the
valence band. Due to the rigid behavior of the BS, the scattering
effect related to the alloying must be quite small.} 
\label{BS2}
\end{figure}

In Fig.\ \ref{BS2} we compare the BS of pure LaRu$_4$Sb$_{12}$ and
La(Ru$_{.75}$Rh$_{.25}$)$_4$Sb$_{12}$ showing that, after alloying, an
indirect gap semiconductor ($E^{ind}_g = $ 0.16 eV) is recovered without
great qualitative changes. The rigid behavior of the BS indicates only weak 
electron scattering due to alloying and the potential for high
mobility. The high effective mass multi-degenerate 
minima in conduction band (Tab.\ \ref{TAB1}) underline the
high Seebeck coefficient that we have calculated from the BS.

Starting from LAPW BS we derived, using
standard kinetic theory,\cite{z-pts,n-etcs} the temperature dependence
of the thermopower at different doping levels giving a direct way
to compare with experimental measures. The transport coefficients are
calculated using (atomic units):
\begin{eqnarray}
{{\sigma}\over{\tau}} &=& {{e^2}\over{3}} \int d\varepsilon (-{{\partial
f_0}\over{\partial \varepsilon}}) N(\varepsilon)v^2(\varepsilon), \\
S &=& {{e \tau}\over{3 \sigma T}} \int d\varepsilon (-{{\partial
f_0}\over{\partial \varepsilon}})
N(\varepsilon)v^2(\varepsilon)(\varepsilon - \epsilon_F), 
\end{eqnarray} 
where $\epsilon_F$ is the chemical potential, $e$ the electron charge, $f_0$
the Fermi distribution function and $v(\varepsilon)$ the average
velocity of electron with energy $\varepsilon$; the constant
scattering time ($\tau$) approximation is assumed.

\begin{figure}%%%%%%%%%%% fig 3
\epsfbox{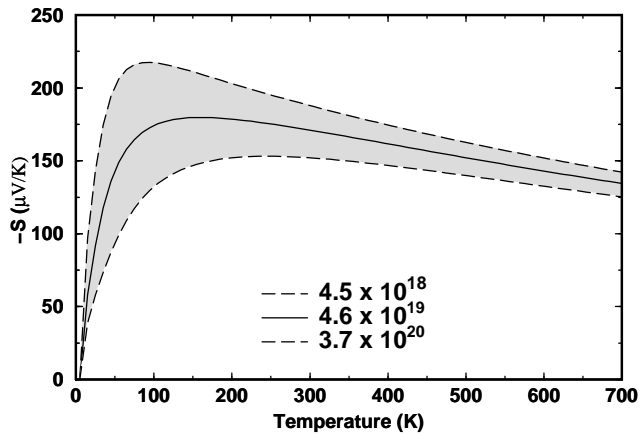}
\caption{Temperature dependence of Seebeck coefficient for
La(Ru$_{.75}$Rh$_{.25}$)$_4$Sb$_{12}$ at different doping
concentrations in electron per cm$^{-3}$.}
\label{S}
\end{figure}

The measured values of Seebeck coefficient for LaFe$_3$CoSb$_{12}$ and
CeFe$_4$Sb$_{12}$ are about 100 $\mu V K^{-1}$ at room
temperature;\cite{smw-s,fbcmm-ict} in the compound we present,
La(Ru$_{.75}$Rh$_{.25}$)$_4$Sb$_{12}$, $S$ ranges from -150 to -200 $\mu V
K^{-1}$ for different doping from 4.5 10$^{18}$ to 3.7 10$^{20}$
electron cm$^{-3}$. As shown in Fig.\ \ref{S} the temperature
dependence is very favorable having a maximum 
below 200 K and remaining quite stable at higher temperature.

In summary, we have presented band structure calculations for RhSb$_3$,
LaRu$_4$Sb$_{12}$ and La(Ru$_{.75}$Rh$_{.25}$)$_4$Sb$_{12}$. Our results
and the experimental trends suggest
La(Ru$_{1-x}$Rh$_{x}$)$_4$Sb$_{12}$ 
as a very promising material for n-type TE application
because of (1) expected low thermal conductivity, (2) multi-valley
high effective mass character of the CB, (3) rigid behavior of the
band structure after alloying, (4) high Seebeck coefficients in the
temperature range useful to cooler device applications. 
It should be emphasized that although the results indicate weak n-type
carrier scattering due to Ru-Rh disorder, the sensitivity of the
band structure to La filling indicates strong scattering by La
defects, so preparing samples with maximum filling is key to
assessing the TE potential of La(Ru$_{1-x}$Rh$_{x}$)$_4$Sb$_{12}$. 
We underline that the difficulty to prepare n-type highly filled
skutterudites has been overcame using high pressure
synthesis,\cite{tmmse-jac} for CoSb$_3$ and this should work also here.

We are grateful to Dr. F.J. Di Salvo for the careful reading and
comments on our manuscript. This work was supported by ONR and DARPA. 

\begin{table}%%%%%%%%%%% tab 1
\caption[TAB. 1]{Effective masses for VCA La(Ru$_{.75}$Rh$_{.25}$)$_4$Sb$_{12}$
along the high symmetry directions in BZ at $\Gamma$, $N$ and at the
minimum near $\Delta$ direction, {\bf K}$^c_m$=(0.170, 0.486,
0.001) in unit of 2$\pi$/a. $\varepsilon$ denotes the band energy with
respect to the top of the valence band (see Fig.\ref{BS2}); $\eta$ is
the number  of equivalent valleys. The effective masses are along
$\Delta$, $\Lambda$, $\Sigma$ for $\Gamma^v$; along $\Sigma$, $G$ and
$D$ for $N^c$. For {\bf K}$^c_m$ the effective masses are along the
cubic directions. Both the six $N^c$ and 24 {\bf K}$^c_m$ pockets are active 
in transport above 150 K for any n-type doping.}

\begin{tabular}{c c c c c c} 
    & $\varepsilon$ (eV)& $\eta$ & $m_1^{\star}$ & $m_2^{\star}$ &
$m_3^{\star}$   \\ \hline
$\Gamma^v$   & 0.000 & 1  & -0.13 & -0.13 & -0.12 \\ \hline
$N^c$        & 0.162 & 6  &  1.16 &  0.36 &  2.00 \\ \hline 
{\bf K}$_m^c$& 0.172 & 24 & 12.12 &  0.24 &  3.90 \\ 
 \end{tabular}
\label{TAB1}
\end{table}

\end{document}